\documentclass[lettersize,journal]{IEEEtran}
\usepackage{amsmath,amsfonts}
\usepackage{algorithmic}
\usepackage{algorithm}
\usepackage{array}
\usepackage[caption=false,font=normalsize,labelfont=sf,textfont=sf]{subfig}
\usepackage{textcomp}
\usepackage{stfloats}
\usepackage{url}
\usepackage{verbatim}
\usepackage{graphicx}
\usepackage{cite}
\usepackage[utf8]{inputenc}
\usepackage[switch, modulo]{lineno}
\usepackage{hyperref}

\begin{document}

\title{A fast plastic scintillator for low intensity proton beam monitoring}

\author{A. André, C. Hoarau, Y. Boursier, A. Cherni, M. Dupont, L. Gallin Martel, M.-L. Gallin Martel, A. Garnier, J. H\'{e}rault, J.-P. Hofverberg, P. Kavrigin, C. Morel, J.-F. Muraz, M. Pinson, G. Tripodo, D. Maneval and S. Marcatili
  % <-this % stops a space
% \thanks{This work was partially supported by INSERM Cancer (TIARA project) and by the European Union (ERC project PGTI, grant number 101040381). Views and opinions expressed are however those of the authors only and do not necessarily reflect those of the European Union or the European Research Council Executive Agency. Neither the European Union nor the granting authority can be held responsible for them.}% <-this % stops a space
\thanks{This work did not involve human subjects or animals in its research.}
\thanks{A. André, C. Hoarau, L. Gallin Martel, M.-L. Gallin Martel, M. Pinson, P. Kavrigin, J.-F. Muraz and S. Marcatili are with Univ. Grenoble Alpes, CNRS, Grenoble INP, LPSC-IN2P3, 38000 Grenoble, France (e-mail: adelie.andre@lpsc.in2p3.fr, sara.marcatili@lpsc.in2p3.fr). }
\thanks{ Y. Boursier, A. Cherni, M. Dupont, A. Garnier and C. Morel are with Aix-Marseille Univ, CNRS/IN2P3, CPPM, Marseille, France.}
\thanks{J. H\'{e}rault, J.-P. Hofverberg and D. Maneval are with Centre Antoine Lacassagne, 06200 Nice, France.}
\thanks{G. Tripodo is with Dipartimento di Fisica e Chimica "E. Segrè", Università degli Studi di Palermo, via delle Scienze, 90128 Palermo, Italy.}
}

%\markboth{Journal of \LaTeX\ Class Files,~Vol.~14, No.~8, August~2024}%

%\IEEEpubid{0000--0000/00\$00.00~\copyright~2024 IEEE}

\maketitle

\begin{abstract}
In the context of particle therapy monitoring, we are developing a gamma-ray detector to determine the ion range \textit{in vivo} from the measurement of particle time-of-flight. For this application, a beam monitor capable to 
tag in time the incident ion with a time resolution below 235~ps FWHM (100~ps rms) is required to provide a start signal for the acquisition. 
We have therefore developed a dedicated detector based on a fast organic scintillator (EJ-204) of 25~$\times$25~$\times$1~mm$^3$ coupled to four SiPM strips that allow measuring the particle incident position by scintillation light sharing. The prototype was characterised with single protons of energies between 63 and 225~MeV at the MEDICYC and ProteusONE facilities of the Antoine Lacassagne proton therapy centre in Nice. 
We obtained a time resolution of 120~ps FWHM at 63~MeV, and a spatial resolution of $\sim$2~mm rms for single particles. Two identical detectors also allowed to measure the MEDICYC proton energy with 0.3\% accuracy.
\end{abstract}


\section{Introduction}
\IEEEPARstart{C}{ompared} to conventional X-ray radiotherapy, Particle Therapy (PT) provides a high ballistic precision with limited irradiation of healthy tissue, thanks to a sharp maximum of the dose deposition at the end of the particle range (Bragg peak). 
However, PT accuracy may be limited by numerous sources of uncertainties such as patient mispositioning, transient modifications of the anatomy and errors in the determination of tissue stopping powers from X-ray images. In order to reduce the safety margins imposed by these uncertainties and improve the technique specificity, particle range has to be measured on-line with millimetric precision\cite{paush_2020}. 
Several groups worldwide are developing different techniques for PT monitoring, taking advantage of the spatial and/or temporal correlation between the dose and the emission vertices of secondary particles like Prompt Gammas (PGs)\cite{KRIMMER_2018}. 
In this context, we are developing a PG detection system that exploits the PG Timing (PGT) technique to determine the ion range from an exclusive measurement of particles Time-Of-Flight (TOF)\cite{hueso_gonzalez_first_2015}. The system consists of a multi-channel PG detectors array (TIARA for Tof Imaging ARrRay) arranged around the target/patient to measure the PG time of arrival $t_{stop}$, which are read-out in coincidence with a fast beam monitor that tags in time the incident ions $t_{start}$. The overall TOF measured ($TOF=t_{stop}-t_{start}$) is equal to the ion transit time in the patient up to the PG vertex, plus the PG TOF from the vertex to the gamma detector. The PG vertex coordinates and subsequently the ion range in the patient can be retrieved from direct measurements of the TOFs as a solution of a non-trivial inverse problem \cite{Jacquet_2021}. For the PGT technique, the higher the system time resolution, the higher the accuracy achieved on the determination of the particle range.\\
We have already demonstrated that a Coincidence Time Resolution (CTR) between the beam monitor and TIARA of 235~ps FWHM allows a proton range accuracy of a few millimeters with a limited statistics of $\sim$10$^7$~protons, equivalent to an average irradiation spot in PT treatments\cite{Jacquet_2021, Jacquet_2023}.
Here we describe the development and characterisation of the beam monitor we have conceived for this application. \\
The monitor must have a time resolution well below the 235~ps FWHM targeted for the whole system CTR for single ions; a detection surface large enough to cover the section of a typical pencil beam (more than 1~cm FWHM at lower proton energies\cite{Pidikiti_2017}); a nearly perfect detection efficiency; a good energy resolution to count incident protons (at least at low intensities). Moreover it must be capable of measuring the beam incident position with an accuracy of the order of the millimeter. 
Several proton beam monitors of this kind are under development. They are based on Ultra Fast
Silicon Detectors (UFSD) (1.2~$\times$~1.2~mm$^{2}$ and 50~µm thick) \cite{Sacchi_2020}, small plastic scintillators (3~$\times$~3~$\times$~3~mm$^{3}$) coupled to Silicon Photomultipliers (SiPM) by optical fibres \cite{GarciaDiez_2024} and diamonds (9~$\times$~9~$\times$~0.5~mm$^{3}$) \cite{Curtoni_2021, Gallin-Martel_2018}, but their limited size is not adapted to clinical beams. A large-area, high time resolution counter based on a plastic scintillator (60~$\times$~30~$\times$~5~mm$^{3}$) read-out by SiPMs was proposed, instead by Cattaneo et al.\cite{Cattaneo_2014}: they achieved a time resolution of 42~$\pm$~2~ps rms with electrons, which approximately deposit 1~MeV in the detector. \\
The detector we have developed is also based on a fast plastic scintillator coupled to SiPMs and is position-sensitive. Its design is described in section~\ref{det_des}.
It was recently tested on the two accelerators of the Centre Antoine Lacassagne (CAL) in Nice: the MEDICYC cyclotron provides 63~MeV protons within a bunch of 4~ns width delivered every 40~ns \cite{Hofverberg_2022} and the IBA synchrocyclotron S2C2 (ProteusONE) provides protons with an energy ranging from 100~to 225~MeV within bunches of 16~ns period and 50\% duty cycle \cite{Pidikiti_2017}. 
Given the pulsed time structure of CAL accelerators, the overall CTR is a convolution of the intrinsic system CTR and the proton bunch width. 
Actually, at clinical intensities, where thousands of protons may be delivered within each bunch, it is impossible to identify the true proton-gamma coincidences and only the proton bunch can be tagged in time. Therefore, in order to fully exploit the potential of our detector, we propose to reduce the beam current to $\sim$1~proton/bunch during the first irradiation spot(s) \cite{Dauvergne_2020}. This approach will allow to verify patient positioning as well as to assess the absence of anatomical modifications occurring before the treatment session with the highest accuracy. 
Preliminary experiments carried out at cyclotron, synchrotron and synchro-cyclotron facilities have shown that the implementation of this regime presents no technical barrier, but the use of a dedicated, high sensitivity beam detector as the one proposed is required to monitor the delivered dose and cope with beam intensity fluctuations at low fluxes.
Notwithstanding, the TOF measurement of single protons of 100-200~MeV are especially challenging because of their limited ionisation capabilities.\\
In this work, the beam monitor performance is thus evaluated at low beam intensities in terms of Detector Time Resolution (DTR), detection efficiency and spatial resolution (section~\ref{charac}). Two identical prototypes were also used to determine the beam energy from particle TOF\cite{Vignati_2020} (section~\ref{e_tof}).
\section{Detector description}\label{det_des}
\begin{figure}[!h]
    \centering
    \includegraphics[width = 5 cm]{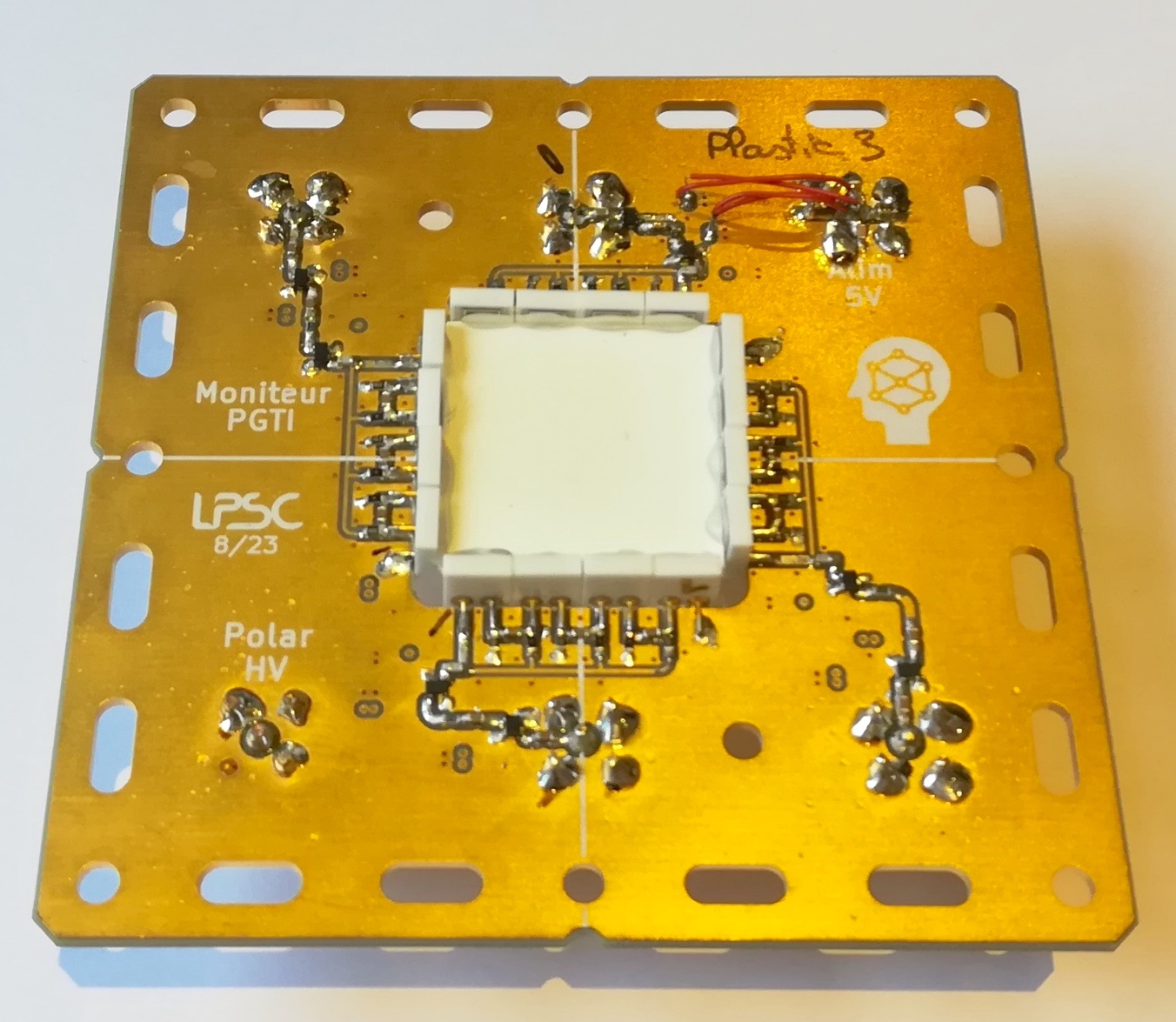} 
    \caption{The beam monitor consists of a 25~$\times$~25~$\times$~1~mm$^{3}$ EJ-204 plastic scintillator read-out by four strips including four SiPMs (3~$\times$~3~mm$^{2}$) each. Signals from SiPMs belonging to the same strip are multiplexed resulting in four electronic readout channels that are separately amplified.}
    \label{fig1}
\end{figure}

\begin{figure}[!h]
    \centering
    \includegraphics[width = 8 cm]{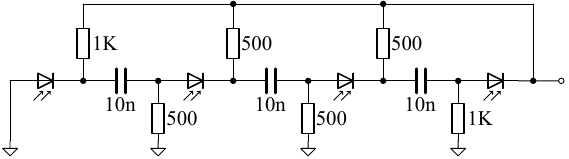}
    \caption{Schematic of the hybrid layout implemented for the readout of each SiPM strip.}
    \label{fig2}
\end{figure}

\begin{figure}[!h]
    \centering
    \includegraphics[width = 7 cm]{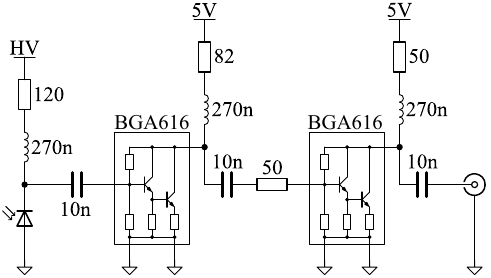}
    \caption{Electrical schematic of the amplifier.}
    \label{fig3}
\end{figure}

\begin{figure}[!h]
    \centering
    \includegraphics[width = 8 cm]{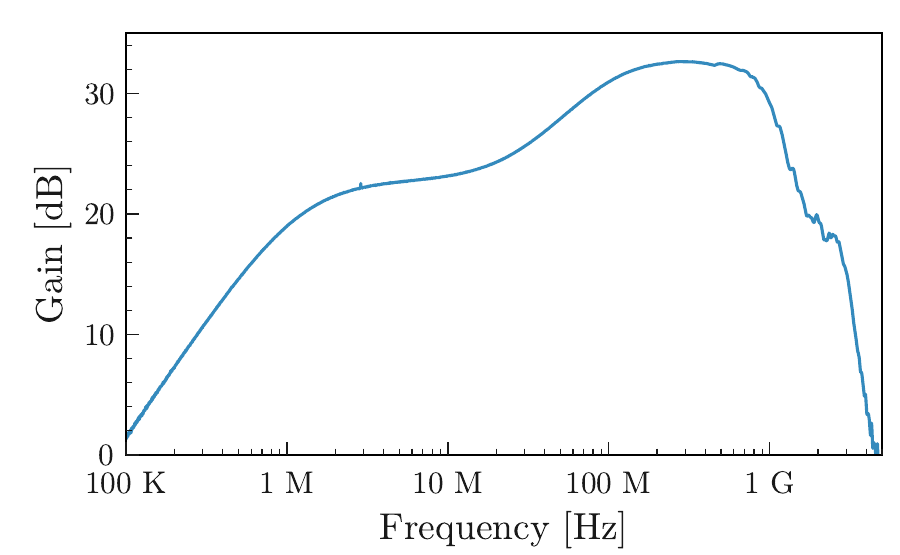}
    \caption{Gain of the amplifier.}
    \label{fig4}
\end{figure}

The beam monitor is composed of a 25~$\times$~25~$\times$~1~mm$^{3}$ fast plastic scintillator (Eljen Technology EJ-204). The two detection surfaces are painted with the EJ-510 reflecting coating while the four edges are coupled to 16~SiPMs (3~$\times$~3~mm$^{2}$ Hamamatsu S13360-3075CS) arranged in four strips as illustrated in Fig.~\ref{fig1}. 
 SiPMs in each strip are mounted with an hybrid layout\cite{Cervi_2017, Nies_2018}: 
the output signals are connected in series, in order to minimise the overall detector capacitance and optimise the time resolution, and the bias voltages are provided through a parallel connection. AC and DC are separated by capacitors (Fig. \ref{fig2}). 
The signal from each SiPM strip goes through a low-noise, RF, two-stages amplifier consisting in heterojunction bipolar integrated circuits, BGA616 from Infeneon (Fig. \ref{fig3}).
The series capacitors are adjusted to control the bandwidth, specifically the low frequency cut-off, so to minimise the noise (Fig. \ref{fig4}).

The signals from the four SiPM strips are acquired separately using the WaveCatcher \cite{Breton_2014} digital sampler, with 500~MHz bandwidth and a sampling rate of 3.5~Gs/s. Data analysis is performed offline in this study. We are currently developing a dedicated data acquisition system based on a fully-digital TDC implemented in the FPGA for the measurement of detector time stamps, coupled to an ADC to measure signals' amplitude for position reconstruction. This approach is possible as the use of ASICs can be avoided thanks to the low number of electronic channels, and it will eventually allow to obtain the detector response in real-time.
\section{Prototype Characterisation}\label{charac}
\subsection{Time resolution}\label{t_charac}
\begin{figure}[!h]
    \centering
    \includegraphics[width = 8 cm]{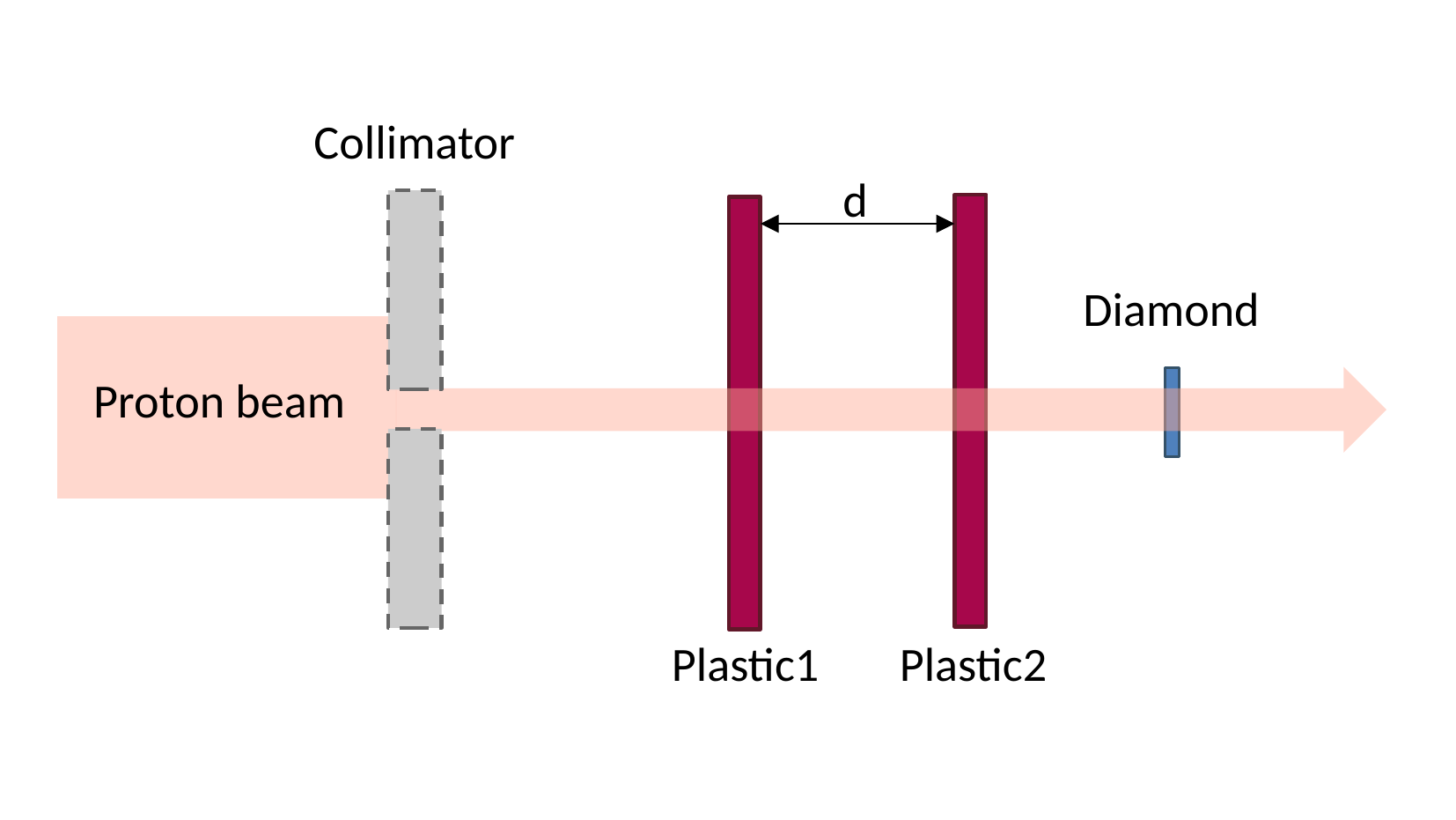}
    \caption{Set-up used for the different characterizations. Two identical prototypes are placed in the beamline to measure their CTR. A diamond placed downstream is used to evaluate detection efficiency. A collimator is added for the spatial resolution characterisation at MEDICYC. 
    The distance $d$ between the two monitors varies from 10 to~95~cm for the TOF-based energy measurement experiment.}
    \label{fig5}
\end{figure}

\begin{figure}[!h]
    \centering
    \includegraphics[width = 8 cm]{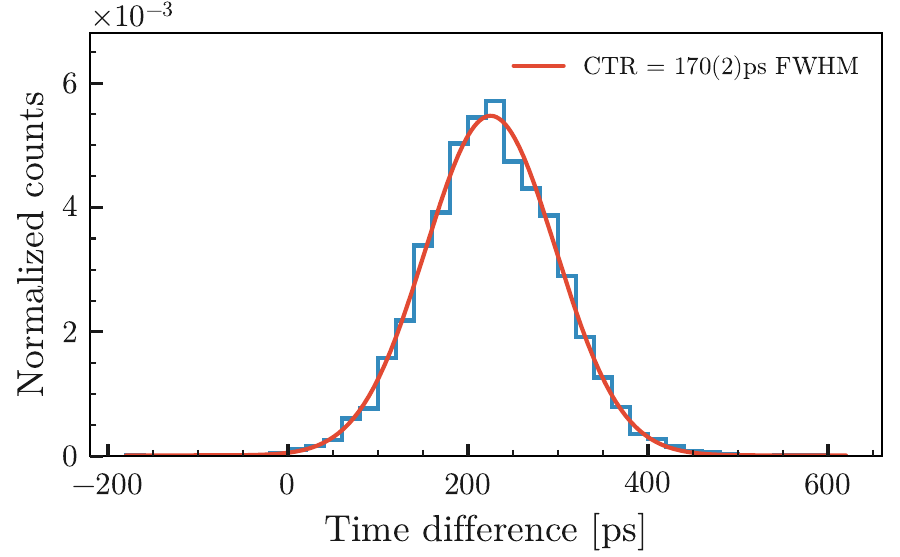}
    \caption{The CTR obtained between two identical beam monitor prototypes is of 170~ps FWHM for 63~MeV protons.}
    \label{fig6}
\end{figure}

\begin{figure}[!h]
    \centering
    \includegraphics[width = 8 cm]{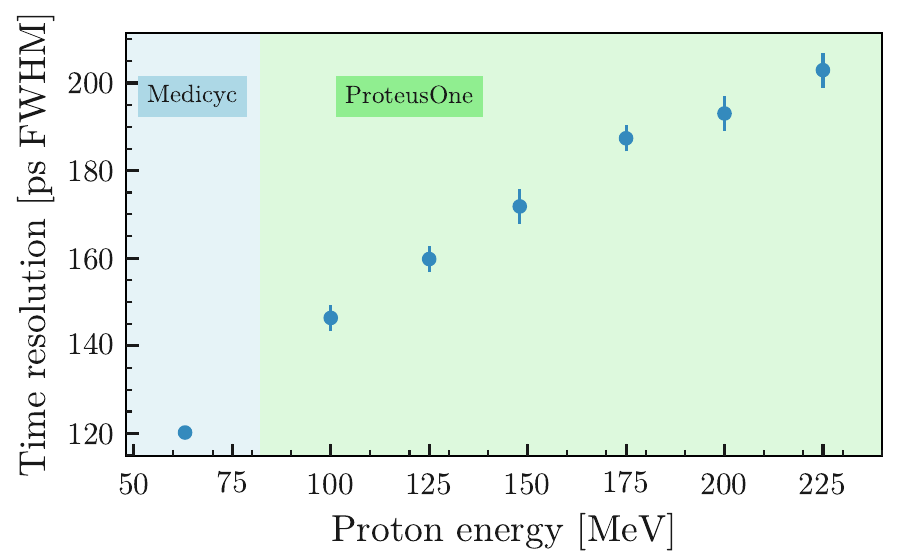}
    \caption{Beam monitor DTR measured at different energies. The value at 63~MeV was obtained at MEDICYC, the other ones at ProteusONE. The DTR increases with proton energy but remains lower than 235~ps FWHM for proton energies within the clinical energy range. }
    \label{fig7}
\end{figure}

The monitor time resolution is determined from the time difference between two identical prototypes placed in the proton beam (Fig. \ref{fig5}). For this measurement, the intensity is reduced in order to detect one proton at a time. For each detector, the particle time is measured as the average of the time stamps obtained with the four strips separately. The latter are determined by implementing a digital Constant Fraction Discriminator (CFD) at 10$\%$ of the maximum amplitude.
The Full Width at Half Maximum (FWHM) of the Gaussian distribution obtained corresponds to the CTR between the two prototypes. Assuming that the two monitors are identical and that protons deposit the same energy in both detectors (protons with an energy comprised between 60 and 225~MeV are almost at the minimum of ionisation), the single DTR is given by: 
\begin{equation}
\label{eq1}
DTR = \frac{CTR}{\sqrt{2}}
\end{equation}
The CTR obtained at the MEDICYC accelerator for 63~MeV protons is of 170~ps FWHM (Fig. \ref{fig6}) corresponding to a DTR of 120~ps FWHM. The same measurement was performed at the ProteusONE synchro-cyclotron for energies between 100 and 225~MeV. Results are plotted in Fig. \ref{fig7}. As expected, the time resolution deteriorates with increasing proton energy, since the energy deposited in the monitor decreases. Nevertheless, in the clinical relevant energy range, i.e., up to 225~MeV, the time resolution remains below the 235~ps FWHM, value which is required for PGT-based range monitoring with TIARA.
\subsection{Detection efficiency}
A diamond detector was placed downstream of the two plastic monitors and used as a trigger. Since diamond detection efficiency was previously determined to be $100\%$ \cite{Curtoni_2021}, the beam monitor efficiency can be measured as the fraction of events detected by the plastic monitor over those detected by the diamond detector. A $100\%$ detection efficiency was obtained on both accelerators at their typical operating energies (corresponding to 63~MeV for MEDICYC and 148~MeV for ProteusONE). Despite this is generally the case for charged particle detectors, this result was not obvious considering the relatively low light yield of the plastic scintillator employed and that the SiPMs only partially cover the scintillator readout surfaces. At the highest relevant energy (225~MeV), where each proton approximately deposits 25\% less energy in the detector, the detection efficiency may be slightly degraded. 
\subsection{Spatial resolution}
The prototype we have developed allows for the determination of incident position of particles within the scintillator surface by measuring the scintillation light shared among opposite SiPM strips. Correcting for the scintillator attenuation length $\lambda$, a parameter proportional to the hit position $x$ is calculated as follows:
\begin{equation}
\label{eq2}
\ln(Q_{1}/Q_{2})=-\frac{2}{\lambda}x +\frac{L}{\lambda}
\end{equation}
where $Q_{1}$ and $Q_{2}$ are the signals' integrals collected by two SiPM strips on opposite sides, and $L$ is the scintillator's width. 
This formula, however, assumes that each strip presents the same light collection efficiency and electronic gain, and cannot be used to estimate the particle hit position, directly. 
Instead, the spatial response was calibrated by gradually moving the prototype from $-$7.5 to 7.5~mm along X and Y axis perpendicular to the beam direction, in steps of 2.5~mm. For each position, 5000~events were acquired to fit a Gaussian distribution centred on the beam position with a standard-deviation $\sigma_{exp}$ that results from the convolution of the actual beam size $\sigma_{beam}$, by the spatial resolution of the beam monitor $\sigma_{det}$: 
\begin{equation}
\label{eq3}
\sigma_{exp}^{2}= \sqrt{\sigma_{det}^2 + \sigma_{beam}^{2}} .
\end{equation}

\begin{figure}[!h]
    \centering
    \includegraphics[width = 8 cm]{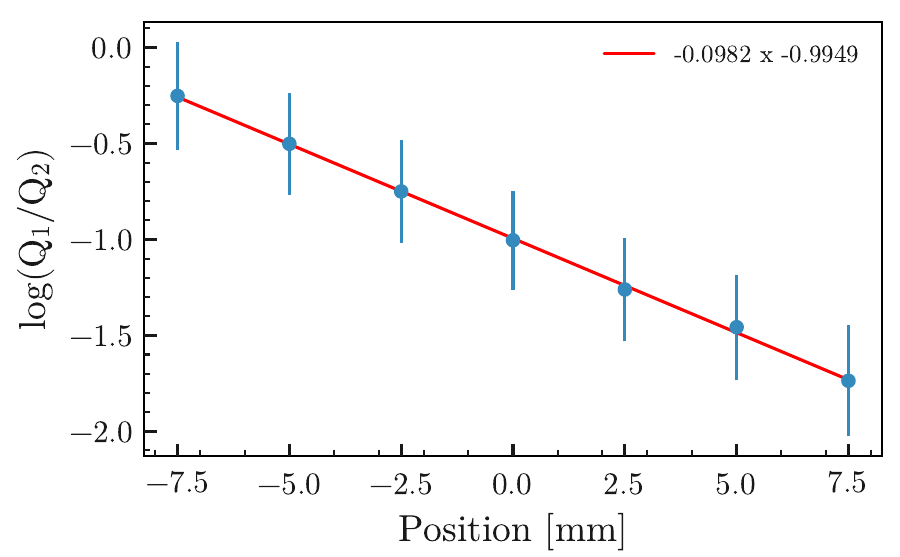}
    \caption{Spatial calibration of the beam monitor. Data obtained with 63~MeV protons at MEDICYC show a linear correlation between the log ratio of the signals' integrals collected by opposite SiPM strips and the actual position of the beam within the scintillator surface.}
    \label{fig8}
\end{figure}

\begin{figure}[!h]
    \centering
    \includegraphics[width = 9 cm]{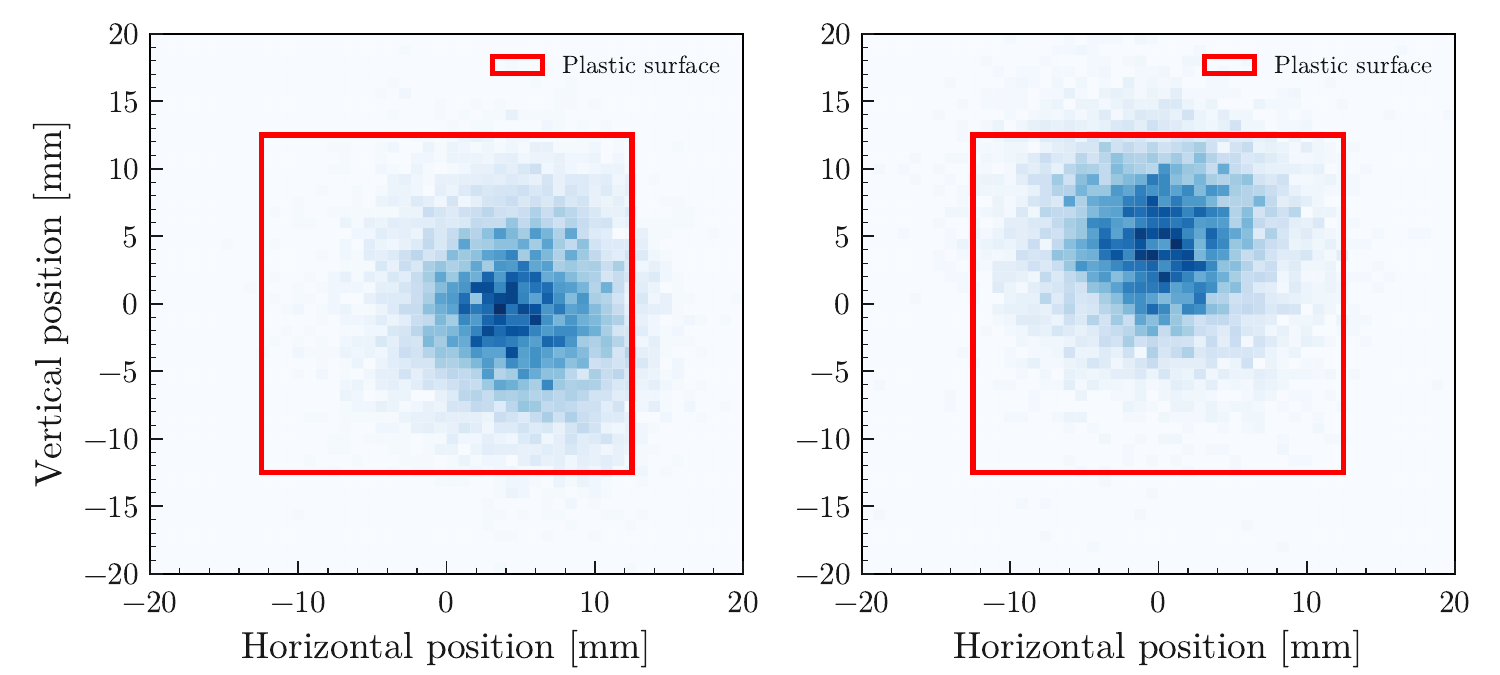}
    \caption{Images obtained at the ProteusONE accelerator with the proton beam shifted in two different positions perpendicularly to the beam direction: (5,~0)~mm on the left and (0, 5)~mm on the right. The red squares represent the monitor detection surface.}
    \label{fig9}
\end{figure}

The beam positions obtained with the 63~MeV protons
at the MEDICYC, are plotted in Fig.~\ref{fig8} as a function of the actual beam monitor displacement.
The calculation of $\ln(Q_{1}/Q_{2})$ permits to obtain a linear response\cite{SWEANY2019} for the detector spatial calibration. The overall spatial resolution is evaluated by converting the width (at 1$\sigma$) of the $\ln(Q_{1}/Q_{2})$ distributions into spatial coordinates, according to the linear fit shown in Fig.~\ref{fig8}. The detector spatial resolution $\sigma_{det}$ is obtained after deconvolution of the beam size. For the latter, values of $1.9~\pm~0.2$~mm rms and $4.3~\pm~0.1$~mm rms are independently measured with a radio-sensitive Gafchromic film, for the 63~MeV protons at MEDICYC and the 148~MeV protons at ProteusONE, respectively. 
With this approach, the detector spatial resolution is estimated at $1.8~\pm~0.3$~mm rms and $2.0 \pm 0.2$~mm rms with the 63~and the 148~MeV protons respectively. These values correspond to the uncertainty in the determination of the incident position of single protons. In the context of our application, however, only the average beam position is needed; since the uncertainty of this parameter decreases as 1/$\sqrt{N}$, with $N$ the number of protons detected, a sub-millimetric accuracy is expected for the average irradiation spot ($\sim$~10$^7$~protons). %
As an example, Fig.~\ref{fig9} shows the 2D images obtained with the beam monitor for the 148~MeV protons shifted by 5~mm in the two directions perpendiculary to the beam. 
\section{TOF-based energy measurement}\label{e_tof}
\begin{figure}[!h]
    \centering
    \includegraphics[width = 8 cm]{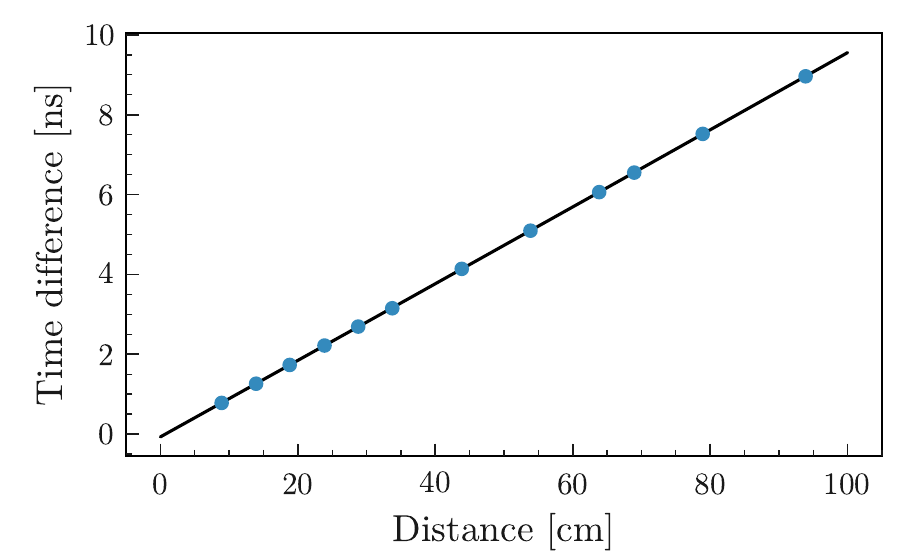}
    \caption{Time difference between the two plastic monitors as a function of their relative distance. Experimentals errors are within the marker size. A $2^{nd}$~degree polynomial fit is used to determine the initial proton speed $v_0$. }
    \label{fig10}
\end{figure}

The two prototypes we have developed have also been used to measure the beam energy at the MEDICYC facility.
The proton kinetic energy depends on the proton speed which can be evaluated by measuring the TOF between two fast detectors separated by a known distance $d$ (cf. Fig.~\ref{fig5}) \cite{Vignati_2020, Galizzi_2018}. 
According to Monte Carlo simulations, for kinetic energies between 60 and 65~MeV, the speed $v$ of a proton travelling through 1~m of air can be considered linear with the position $x$: 
\begin{equation}
\label{eq4}
v(x) = a\times x + v_0
\end{equation}
with $a$ the proton velocity loss per unit distance and $v_0$ the proton initial speed (at the exit of the upstream monitor). Thus, the time difference $TOF_{exp}$ measured between the two monitors can be written as the actual proton TOF plus a constant, $c$, that models the systematic delay induced by cables and electronics:
\begin{equation}
\label{eq5}
TOF_{exp} = \frac{\log (\frac{a}{v_0} x + 1)}{a} + c
\end{equation}
With $\frac{a}{v_0} x <$0.015~ns for x$<$100~cm, the previous equation can be approximated by the following Taylor series:
\begin{equation}
\label{eq6}
TOF_{exp} = -\frac{a}{2 v_0^{2}}x^{2}+\frac{1}{v_0}x+c + \mathcal{O}(x^{3})
\end{equation}
where $\mathcal{O}(x^{3})<10^{-3}$~ns for $x\leq$~100~cm (calculated from the fit parameters) and it is therefore negligible. \\The time difference between two monitors was measured for several distances by moving forwards the downstream monitor (Plastic2 on Fig. \ref{fig5}). The results obtained are plotted in Fig.~\ref{fig10} and fitted with equation~\ref{eq6} to determine the proton speed $v_0$ at the exit of the upstream monitor (Plastic1). From this value, the proton kinetic energy $E_0$ can be calculated as:
\begin{equation}
\label{eq7}
E_0 = m c^{2} (\gamma - 1) 
\end{equation}
with $m$ the proton mass, $c$ the speed of light and $\gamma~=~1/\sqrt{1-\beta ^{2}}$ with $\beta=v_0/c$. A value of $62.52\pm 0.18$~MeV was obtained for the MEDICYC cyclotron.
Knowing the composition of the scintillator, it is then possible to correct for the energy loss in the upstream beam monitor. In this study, the energy loss was evaluated by Monte Carlo simulation with an uncertainty negligible compared to the experimental one, and allowed to establish an initial kinetic energy of 63.58~$\pm$~0.18~MeV. For comparison, at this energy, $0.18$~MeV corresponds to a proton range error of 0.17~mm in liquid water.

Furthermore, the parameter $a$ in equation~\ref{eq4} gives the mean speed loss per unit distance and was estimated at $-$1.0~$\pm$~0.4~keV/mm. The accuracy of this measurement is low but the result is consistent with the NIST PSTAR value of $-$1.1~keV/mm \cite{nist}.\\

The proposed method allows to disregard systematic errors made in the measurement of proton $TOF_{exp}$ (due to cable delays and electronics) and the real distance between the monitors (only relative displacements $d$ must be precisely measured). The relative error on $E_0$ derived from equation \ref{eq7} is given by:
\begin{equation}
\label{eq8}
\frac{\Delta E_0}{E_0}= \frac{\beta^2 \gamma^3}{\gamma -1 }  \times\frac{\Delta v_0}{v_0}.
\end{equation}

With repeated measurements at different distances, we were able to reduce the overall error on energy but still, the single measurement is affected by the resolution of the instruments employed. 
Considering that the second order of the expansion given by equation \ref{eq6} is negligible ($\frac{a}{2 v_{0}^{2}} \cong 0.7\times10^{-3}~ns/cm \ll \frac{1}{v_0}\cong0.1~ns/cm$), and assuming that the time and the distance are measured independently, the relative uncertainty on $v_0$ can be estimated by the quadratic contributions of the relative uncertainties on the measurement of time and distance:
\begin{equation}
\label{eq9}
\frac{\Delta E_0}{E_0}=\left( \frac{\beta^2 \gamma^3}{\gamma -1 } \right) \sqrt{\left( \frac{\Delta TOF}{TOF}\right)^{2} +\left( \frac{\Delta d}{d} \right)^{2}}
\end{equation}
Here, $\Delta TOF$ is the CTR between the two monitors, equal to 170~ps FWHM at the energy of the MEDICYC cyclotron (cf. section~\ref{t_charac}), divided by the square root of the number of protons measured for each position (5000), and $\Delta d$ is the error made when measuring the downstream monitor displacement. With our detector, the first term is negligible since we are measuring TOFs of the order of the nanosecond, while the second term completely dominates the overall accuracy as the  distance was measured with an external caliber and was therefore affected by parallax error. In the future, it will be possible to improve the sensitivity of the technique by placing the two detectors on a linear stage with sub-millimeter accuracy. 
\section{Conclusion}
We developed a large-area, fast, position-sensitive beam monitor based on a plastic scintillator read-out by SiPMs. This device will be used as a start trigger for the TIARA detector\cite{Jacquet_2021, Jacquet_2023} with the goal of measuring the ion range \textit{in vivo}, with the PGT technique. 
The detector was characterised with single protons at the MEDICYC and ProteusONE facilities (CAL, Nice). 
A time resolution below 235~ps FWHM, as required for PGT measurement, was obtained within the clinically relevant energy range, i.e., up to 225~MeV, with a best value of 120~ps FWHM for single protons of 63~MeV. For higher beam intensities the monitor time resolution is expected to improve, but this is of no advantage for the determination of proton-gamma coincidences in our application. 
Thanks to the excellent time performances achieved at 63~MeV, it was possible to measure the proton energy at the MEDICYC cyclotron with a relative accuracy of 0.3\% (0.18~MeV). The main source of uncertainty for this experiment, was the determination of the relative distance between the two monitors: this measurement can be improved in the future by implementing a dedicated linear translation system with sub-millimeter accuracy. 
In addition, the monitor can measure the single-particle incident position with a spatial resolution of $\sim$2~mm rms, and the average beam position with an accuracy that improves with the number of incident protons, therefore ensuring an uncertainty well below the millimeter for a typical irradiation spot in a PT treatment.\\
It should be noted, however, that this detector is probably too radiation-sensitive to withstand the typical annual doses delivered by a clinical irradiation facility. The main issue, in this sense, is the relative proximity of SiPMs to the beam axis. For a future prototype we therefore plan to increase the scintillator surface: this would allow to limit SiPM irradiation and to get a detector large enough to be compatible with pencil beam scanning.
\section*{Acknowledgments}
All authors declare that they have no known conflicts of interest in terms of competing financial interests or personal relationships that could have an influence or are relevant to the work reported in this paper. 

This work was partially supported by INSERM Cancer (TIARA project) and by the European Union (ERC project PGTI, grant number 101040381). Views and opinions expressed are however those of the authors only and do not necessarily reflect those of the European Union or the European Research Council Executive Agency. Neither the European Union nor the granting authority can be held responsible for them.

\bibliographystyle{IEEEtran}

\bibliography{bibliography}

\vfill

\end{document}